\documentstyle[twoside]{article}

\catcode`\@=11
\long\def\@makefntext#1{
\protect\noindent \hbox to 3.2pt {\hskip-.9pt  
$^{{\eightrm\@thefnmark}}$\hfil}#1\hfill}		

\def\@makefnmark{\hbox to 0pt{$^{\@thefnmark}$\hss}}	
	
\def\ps@myheadings{\let\@mkboth\@gobbletwo
\def\@oddhead{\hbox{}
\rightmark\hfil\eightrm\thepage}   
\def\@oddfoot{}\def\@evenhead{\eightrm\thepage\hfil
\leftmark\hbox{}}\def\@evenfoot{}
\def\sectionmark##1{}\def\subsectionmark##1{}}

\oddsidemargin=\evensidemargin
\newcounter{sectionc}\newcounter{subsectionc}\newcounter{subsubsectionc}
\renewcommand{\section}[1] {\vspace{12pt}\addtocounter{sectionc}{1} 
\setcounter{subsectionc}{0}\setcounter{subsubsectionc}{0}\noindent 
	{\tenbf\thesectionc. #1}\par\vspace{5pt}}
\renewcommand{\subsection}[1] {\vspace{12pt}\addtocounter{subsectionc}{1} 
	\setcounter{subsubsectionc}{0}\noindent 
	{\bf\thesectionc.\thesubsectionc. {\kern1pt \bfit #1}}\par\vspace{5pt}}
\renewcommand{\subsubsection}[1] {\vspace{12pt}\addtocounter{subsubsectionc}{1}
	\noindent{\tenrm\thesectionc.\thesubsectionc.\thesubsubsectionc.
	{\kern1pt \tenit #1}}\par\vspace{5pt}}

\newcounter{appendixc}
\newcounter{subappendixc}[appendixc]
\newcounter{subsubappendixc}[subappendixc]
\renewcommand{\thesubappendixc}{\Alph{appendixc}.\arabic{subappendixc}}
\renewcommand{\thesubsubappendixc}
	{\Alph{appendixc}.\arabic{subappendixc}.\arabic{subsubappendixc}}

\renewcommand{\appendix}[1] {\vspace{12pt}
        \refstepcounter{appendixc}
        \setcounter{figure}{0}
        \setcounter{table}{0}
        \setcounter{lemma}{0}
        \setcounter{theorem}{0}
        \setcounter{corollary}{0}
        \setcounter{definition}{0}
        \setcounter{equation}{0}
        \renewcommand{\thefigure}{\Alph{appendixc}.\arabic{figure}}
        \renewcommand{\thetable}{\Alph{appendixc}.\arabic{table}}
        \renewcommand{\theappendixc}{\Alph{appendixc}}
        \renewcommand{\thelemma}{\Alph{appendixc}.\arabic{lemma}}
        \renewcommand{\thetheorem}{\Alph{appendixc}.\arabic{theorem}}
        \renewcommand{\thedefinition}{\Alph{appendixc}.\arabic{definition}}
        \renewcommand{\thecorollary}{\Alph{appendixc}.\arabic{corollary}}
        \renewcommand{\theequation}{\Alph{appendixc}.\arabic{equation}}
        \noindent{\tenbf Appendix \theappendixc #1}\par\vspace{5pt}}
\newcommand{\subappendix}[1] {\vspace{12pt}
        \refstepcounter{subappendixc}
        \noindent{\bf Appendix \thesubappendixc. {\kern1pt \bfit #1}}
	\par\vspace{5pt}}
\newcommand{\subsubappendix}[1] {\vspace{12pt}
        \refstepcounter{subsubappendixc}
        \noindent{\rm Appendix \thesubsubappendixc. {\kern1pt \tenit #1}}
	\par\vspace{5pt}}

\newcommand{\textlineskip}{\baselineskip=13pt}
\newcommand{\smalllineskip}{\baselineskip=10pt}

\def\eightcirc{
\begin{picture}(0,0)
\put(4.4,1.8){\circle{6.5}}
\end{picture}}
\def\eightcopyright{\eightcirc\kern2.7pt\hbox{\eightrm c}} 

\newcommand{\copyrightheading}[1]
	{\vspace*{-2.5cm}\smalllineskip{\flushleft
	{\footnotesize International Journal of Modern Physics A, #1}\\
	{\footnotesize $\eightcopyright$\, World Scientific Publishing
	 Company}\\
	 }}

\def\abstracts#1#2#3{{
	\centering{\begin{minipage}{4.5in}\baselineskip=10pt\footnotesize
	\parindent=0pt #1\par 
	\parindent=15pt #2\par
	\parindent=15pt #3
	\end{minipage}}\par}} 

\renewenvironment{thebibliography}[1]
	{\frenchspacing
	 \ninerm\baselineskip=11pt
	 \begin{list}{\arabic{enumi}.}
	{\usecounter{enumi}\setlength{\parsep}{0pt}
	 \setlength{\leftmargin 12.7pt}{\rightmargin 0pt} 
	 \setlength{\itemsep}{0pt} \settowidth
	{\labelwidth}{#1.}\sloppy}}{\end{list}}

\newcounter{itemlistc}
\newcounter{romanlistc}
\newcounter{alphlistc}
\newcounter{arabiclistc}

\newcommand{\fcaption}[1]{
        \refstepcounter{figure}
        \setbox\@tempboxa = \hbox{\footnotesize Fig.~\thefigure. #1}
        \ifdim \wd\@tempboxa > 5in
           {\begin{center}
        \parbox{5in}{\footnotesize\smalllineskip Fig.~\thefigure. #1}
            \end{center}}
        \else
             {\begin{center}
             {\footnotesize Fig.~\thefigure. #1}
              \end{center}}
        \fi}

\newcommand{\tcaption}[1]{
        \refstepcounter{table}
        \setbox\@tempboxa = \hbox{\footnotesize Table~\thetable. #1}
        \ifdim \wd\@tempboxa > 5in
           {\begin{center}
        \parbox{5in}{\footnotesize\smalllineskip Table~\thetable. #1}
            \end{center}}
        \else
             {\begin{center}
             {\footnotesize Table~\thetable. #1}
              \end{center}}
        \fi}

\def\@citex[#1]#2{\if@filesw\immediate\write\@auxout
	{\string\citation{#2}}\fi
\def\@citea{}\@cite{\@for\@citeb:=#2\do
	{\@citea\def\@citea{,}\@ifundefined
	{b@\@citeb}{{\bf ?}\@warning
	{Citation `\@citeb' on page \thepage \space undefined}}
	{\csname b@\@citeb\endcsname}}}{#1}}

\newif\if@cghi
\def\cite{\@cghitrue\@ifnextchar [{\@tempswatrue
	\@citex}{\@tempswafalse\@citex[]}}
\def\citelow{\@cghifalse\@ifnextchar [{\@tempswatrue
	\@citex}{\@tempswafalse\@citex[]}}
\def\@cite#1#2{{$\null^{#1}$\if@tempswa\typeout
	{IJCGA warning: optional citation argument 
	ignored: `#2'} \fi}}

\def\pmb#1{\setbox0=\hbox{#1}
	\kern-.025em\copy0\kern-\wd0
	\kern.05em\copy0\kern-\wd0
	\kern-.025em\raise.0433em\box0}

\def\fnt#1#2{\footnotetext{\kern-.3em
	{$^{\mbox{\scriptsize #1}}$}{#2}}}

\def\fpage#1{\begingroup
\voffset=.3in
\thispagestyle{empty}\begin{table}[b]\centerline{\footnotesize #1}
	\end{table}\endgroup}

\def\runninghead#1#2{\pagestyle{myheadings}
\markboth{{\protect\footnotesize\it{\quad #1}}\hfill}
{\hfill{\protect\footnotesize\it{#2\quad}}}}
\font\tenrm=cmr10
\font\tenit=cmti10 
\font\tenbf=cmbx10
\font\bfit=cmbxti10 at 10pt
\font\ninerm=cmr9

\font\eightrm=cmr8

\def\qed{\hbox{${\vcenter{\vbox{			
   \hrule height 0.4pt\hbox{\vrule width 0.4pt height 6pt
   \kern5pt\vrule width 0.4pt}\hrule height 0.4pt}}}$}}

\begin{document}

\runninghead{Brane Cosmology with Non-Static Bulk} 
{Brane Cosmology with Non-Static Bulk}

\normalsize\textlineskip
\thispagestyle{empty}
\setcounter{page}{1}

\copyrightheading{}			

\vspace*{0.88truein}

\fpage{1}
\centerline{\bf BRANE COSMOLOGY WITH NON-STATIC BULK}
\vspace*{0.37truein}
\centerline{\footnotesize ALI NAYERI\footnote{
Talk given at DPF 2000:  The meeting of the Division of
Particles and Fields of the American Physical Society,
Ohio, 9 - 12 August 2000.}}
\vspace*{0.015truein}
\centerline{\footnotesize\it Department of Physics, Massachusetts
Institute of Technology}
\baselineskip=10pt
\centerline{\footnotesize\it 77 Massachusetts Avenue,
Cambridge, MA 02135, U.S.A.}


\vspace*{0.21truein}
\abstracts{We study the brane world motion in non-static bulk by
generalizing the second Randall-Sundrum scenario   
Explicitly,
we take the bulk to be a {\it Vaidya-AdS} metric, which
describes the gravitational collapse of a spherically 
symmetric null
dust fluid in Anti-de Sitter spacetime. We point out
that during an inflationary phase on the brane, black holes 
will tend to be thermally nucleated in the bulk
We analyze the thermodynamical properties of  this brane-world.We 
point out
that during an inflationary phase on the brane, black holes will tend
to be thermally nucleated in the bulk.  Thermal equilibrium of the 
system is discussed. We calculate the late time
behavior of this system, including 1-loop effects.  We argue that
at late times a sufficiently large black hole will relax to a point of
thermal equilibrium with the brane-world environment.
This result has interesting implications
for early-universe cosmology.}{}{}

\textlineskip			
\vspace*{12pt}			


Following is the summary of our earlier paper~\cite{aaa}
in which we generalized the second Randall-Sundrum scenario~\cite{RS2}
to have a non-static bulk.

The second Randall-Sundrum model may be understood
by looking at the AdS/CFT~\cite{gubser}
duality in the non-gravity decoupled limit.  Thus one would expect that the Hawking-Page type of phase transition~\cite{hawkpage}
should occur in that setting.  In particular, during an
inflationary phase on the brane, the brane-world is a
de Sitter hyperboloid embedded in $AdS_5$, and it will generate an
acceleration horizon in the bulk.  This horizon will have a temperature,
and so we would expect that inflating brane-worlds would be unstable
to the creation of bulk (five-dimensional) black holes.
This process was discussed
by Garriga and Sasaki \cite{garriga}.  They
studied the `thermal instantons' which correspond to black
holes in AdS, and showed that these instantons describe the
thermal nucleation of Schwarzschild-AdS black holes in the presence of 
a pre-existing ${\bf Z}_2$ symmetric inflating brane-world.
This could be an evidence that we should {\it assume} that a bulk
black hole will be formed during the inflationary epoch of a 
brane-world universe. Here we assume the brane-world
is like a uniformly accelerating mirror: A `black box' with perfectly
reflecting, accelerating walls.  However, it is well known
\cite{bd} that a uniformly accelerating mirror in five dimensions (with 
acceleration parameter $A$) emits at late times a thermal flux of radiation
at a temperature $T ~{\sim}~ A$
and a corresponding renormalized energy density
$
\langle T_{vv}\rangle ~{\sim}~ A^5,
$
where $v$ is the null direction normal to the brane.  Among different
possibilities, the
most interesting case is the 
case that $T_{BW} > T_{BH}$. Clearly, this
case is of interest because it would appear that the black hole
might be unstable to some runaway process where the mass increases
indefinitely.  
In order to study this case, we allow the bulk to be non-static,
so that radiation emitted by the brane-world will generate 
a backreaction and cause the bulk black hole
to grow.  Since we are interested in the late time solution, we will
only consider ingoing radiation. In 1951, Vaidya \cite{vaidya} wrote down a metric that represents an
imploding (or exploding) null dust fluid with spherical symmetry in asymptotically
flat space.  Recently \cite{rio}, this metric has been generalized to 
describe gravitational collapse in spacetimes with a non-vanishing 
cosmological constant.  Here, we are interested in the metric which
describes gravitational collapse in a spacetime which is asymptotically
Anti-de Sitter (AdS).  This metric is written using 
`Eddington-Finkelstein' coordinates, so that it takes the explicit form
$
ds^{2} =  - f(v,r)dv^{2} + 2 dvdr + r^{2}d\Omega^{2}_{3},
$ where
$
f(v,r) = k - \frac{2M(v, r)}{r^2},
$ and $d\Omega^{2}_{3}$ is the metric on 3 - sphere.
The {\it mass function}, $M(v,r)$, in $AdS_5$ , in general, is
$
\frac{\Lambda}{12}r^{4} + m(v) - \frac{q^2(v)}{r^2},
$
where ${\Lambda} = -(6/l^2)$ is the bulk cosmological
constant, and $m(v)$ is an arbitrary function of $v$ which 
will be determined
by the energy density of the radiation in the bulk.  $q(v)$ 
corresponds to
the charge of the bulk, if any.  The equations of motion for a domain 
wall, when the effects of 
gravity are included, are given by the Israel junction conditions.
These conditions relate the discontinuity in the extrinsic curvature 
($K_{A B}$)
at the wall to the energy momentum ($t_{A B}$) of fields which live
on the wall:
$
\left [K_{A B} - Kh_{A B}\right]^{\pm} = \kappa^{2}_{D} t_{A B}.
$
(see \cite{dddw} for a derivation of this equation).
The gravitational coupling constant, $\kappa^{2}_{D}$, in arbitrary 
dimension $D$
is given by~\cite{rman}
$
\kappa^{2}_{D} = \frac{2(D - 2)\pi^{\frac{1}{2}(D - 1)}}
                      {(D - 3)(D/2 - 3/2)!}G_D, 
$
where $G_D$ is the $D$-dimensional Newton constant.  Here 
$\kappa^{2}_{5} = 3 \pi^2 G_5$.  Since we are interested in the
cosmological aspects of brane-world evolution, we want the metric
induced on the brane-world to assume a manifestly FRW form:
$
ds^{2}|_{brane-world} = 
- d\tau^{2} + a^2(\tau) d\Omega^{2}_{3},  
$
where $\tau$ is the
time experienced by observers who move with the brane-world:
$
d\tau = \left(f - 2\frac{da(\tau)}{dv}\right)^{1/2}dv,
$.  We assume that the stress-energy tensor 
describing the fields which propagate in the brane-world is given as
$
t^{A}_{B} = \mathrm{diag}(-(\rho+\rho_{\lambda}),p-\rho_{\lambda},
p-\rho_{\lambda},p-\rho_{\lambda},0)
$. We emphasize that $\rho$ and $p$ are the energy density and pressure
of the ordinary matter, respectively, whereas, $\rho_{\lambda}$ is the 
contribution from the tension of the brane which is simply a Nambo-Goto
term.  From the Israel equations we may then derive the Friedmann 
equation 
on the brane:
\begin{equation}
\label{Friedmann} 
H^{2}(\tau) = {\Lambda_{eff}\over3} - \frac{k}{a^2} +
\left({8\pi G_4\over 3}\right) \rho  
 \left({\pi}^2 G_5\right)^2\rho^2+{{2m(\tau,a)} \over
a^4} - \frac{q^2}{a^6}\,,
\end{equation}

\noindent Henceforth, we shall set $q = 0$ and 
$k$ the (spatial) curvature of the brane to unity and
$G_5$ to $ \sqrt{4G_4/3\pi\rho_{\lambda}}$.  
$\Lambda_{eff} {\sim} -(1/{l_{eff}})^2$ is
the 4-dimensional cosmological constant on the brane, which is given 
in terms
of the brane tension $\rho_{\lambda}$ and the bulk cosmological
constant $\Lambda$: $\Lambda_{eff} \equiv \Lambda_4 = 
\left( \frac{\Lambda}{2} + 4\pi G_4
\rho_{\lambda} \right)$.  Since we are interested in the evolution 
of bulk black
holes which are as large as the AdS length scale $l$, we are 
implicitly assuming
that $l_{eff} >> l$.  Thus, we find that a time-dependent mass in 
the bulk gives
rise to a time-dependent term that scales like radiation.  In other 
words,
the collapse of radiation (to form a black hole) in the bulk gives 
rise to a
component of `Hot Dark Matter' on the brane.  To understand how this 
term will affect brane-world cosmology,
we would like to solve for the back-reaction
generated by 1-loop effects in the bulk. We will simply estimate 
the late-time behavior of the system.
Here, we 
assume that the only matter in the bulk is radiation which
is emitted by the brane-world as it accelerates away from the bulk 
black hole.
In other words, we want to see if it is even {\it consistent} to 
ignore the 
Hawking radiation.
Following \cite{rio}, we may then relate the intensity of the stress-energy
tensor in the bulk  with the rate of change in the mass function:
$
A^5 ~{\sim}~ \langle T_{vv} \rangle = \frac{2{\dot m}(\tau,a)}{{\kappa}^{2}_{5}
{r_h}^2}\frac{f}{\dot{a} + \sqrt{\dot{a}^2 + f}}
$
where $r_h$ denotes the horizon radius.
The acceleration five-vector for a domain wall moving in a spherically 
symmetric background is given as \cite{dddw}:
$
A^{A} = - K_{\tau \tau} n^{A}
$. So the equation for the mass function becomes
\begin{equation}
\frac{2{\dot m}(\tau,a)}{{\kappa}^{2}{r_h}^2} {\sim}
-\frac{1}{f}\left[\frac{d}{da}\left({\sqrt{f + 
{\dot a}^2}}\right) 
- \frac{1}{2}\frac{\dot{f}}{f^2}\left(\dot{a} + \sqrt{f + \dot{a}^2}\right)^2
\right]^5\left(\dot{a} + \sqrt{\dot{a}^2 + f}\right)
\end{equation}
We now turn to the late time evolution of this equation.
Before the black hole forms, the brane is exponentially expanding.
Since we {\it assume} the brane doesn't fall into the black hole
once the hole forms, the brane should still be exponentially expanding
once gravitational collapse begins.  Consequently,
suppose $a ~{\sim}~ e^{t}$ at late times.  Then it is straightforward
to show that $m \sim e^{-2t}$, and so it would seem that the black hole shrinks.
In fact, this means that at late times our approximation 
always breaks down:  At late times 
we will have to include the effects of Hawking radiation.
Eventually, the black hole will emit a temperature comparable to the
temperature of the brane-world, and the system will equilibrate.  From the 
point of view of the AdS/CFT duality, this makes sense.
Truncating $AdS_5$ with a de Sitter brane introduces a definite 
temperature into the system, namely, the temperature of the `mirror'.
As long as sufficiently large bulk black holes are allowed, we know
that these holes can be in thermal equilibrium with thermal radiation.
The main shortcoming of this analysis is that we assumed reflecting boundary conditions 
for all bulk modes at the brane-world.  

{\noindent \bf Acknowledgements}\\
Author is supported by NSF grant ACI-9619019.

\end{document}